\documentclass{elsart}
\usepackage{graphicx}

\def\eqref#1{eq.~(\ref{#1})}
\def\figref#1{Fig.~\ref{#1}}

\begin{document}

\begin{frontmatter}

\title{Zipf Law in Firms Bankruptcy}
\author[ATR]{Yoshi Fujiwara\thanksref{contact}}
\ead{yfujiwar@atr.co.jp}
\address[ATR]{%
  ATR Human Information Science Laboratories,
  Kyoto 619-0288, Japan}
\thanks[contact]{%
  Corresponding author. FAX: +81-774-95-2647.
}

\begin{abstract}
  Using an exhaustive list of Japanese bankruptcy in 1997, we discover
  a Zipf law for the distribution of total liabilities of bankrupted
  firms in high debt range. The life-time of these bankrupted firms
  has exponential distribution in correlation with entry rate of new
  firms. We also show that the debt and size are highly correlated, so
  the Zipf law holds consistently with that for size distribution. In
  attempt to understand ``physics'' of bankruptcy, we show that a
  model of debtor-creditor dynamics of firms and a bank, recently
  proposed by economists, can reproduce these phenomenological
  findings.
\end{abstract}

\begin{keyword}
  Zipf law \sep firm growth \sep bankruptcy
  \sep balance-sheet \sep Econophysics
  \PACS 89.90.+n \sep 02.50.-r \sep 05.40.+j \sep 47.53.+n
\end{keyword}


\end{frontmatter}


\section{Introduction}

It is a challenging task to model aggregate dynamics of firms in
econophysics. As such dynamics for firms with considerable amount of
money flow is concerned, one can now have evidence of several
phenomenological facts about the dynamics of firms
\cite{SIMON77,Stanley96,Amaral97,Buldyrev97,Sutton97,Amaral98,Okuyama99,%
Axtell01,Mizuno02,Gabaix02,Yoshi03,Aoyama03}, which are not only
interesting in physics, but also have quite significance in
macroeconomics \cite{Sutton97,Gabaix02}.

Among those facts are size distribution and growth of firms.
Distribution of firm size such as sales, number of employees,
total assets follows a power-law, or Pareto-Zipf law
\cite{SIMON77,Amaral97,Okuyama99,Axtell01,Aoyama03}. As for the growth
of firms, the distribution of the logarithm of growth rates has an
exponential form and that the fluctuations in the growth rates scale
with the firm size
\cite{Stanley96,Amaral97,Buldyrev97,Amaral98,Okuyama99,Mizuno02}.
While these observation concerns about stock of a firm, no less
important is flow such as profits, whose properties of distribution
and growth rates have recently observed in empirical works
\cite{Yoshi03,Aoyama03}.

While these phenomenological facts have been found for {\it living\/}
firms, little is known about the {\it death\/} in aggregate dynamics
of a large number of firms. That is exactly what the present work
concerns about. It is important to uncover phenomenology of how firms
die in order to understand the firms growth dynamics.

In section 2, we make observation of Japanese bankruptcy by using an
exhaustive list of bankrupted firms in a year. We show
phenomenological findings about the distribution of total liabilities
or debts when bankrupted, the life-time of bankrupted firms, and
correlation between the firm's debt and size. These are the main
results of this paper. In section 3, we describe a firm's financial
activity in terms of balance-sheet dynamics. A model for debtor of
firms and creditor of a bank, recently proposed by economists
\cite{Gallegati03}, after briefly recapitulated, is shown to be able
to reproduce our findings about bankruptcy. Summary and discussion is
given in section 4. In appendix, we give description about the nature
of bankruptcy dataset employed in this paper.

\section{Firms Bankruptcy in Japan}

Our data, provided by a domestic data-bank company (Tokyo Shoko
Research, Ltd.), is an exhaustive list of Japanese bankrupted firms in
1997. The data is exhaustive in the sense that any bankrupted firm
with total amount of debt exceeding 10 million yen (roughly 80,000
euros/dollars) is listed in it. The number of such firms in the year
is 16,526. The data contains information of debt, sales, business
sector, number of employees, primary bank when bankrupted, dates
of establishment and failure, and stockholders fund when established.
Two remarks immediately follow.

First, bankruptcy or business failure is not a legal term, but should
be understood as a critical financial insolvency of a debtor. Typical
case and classification of different types are given in appendix. What
is important for our purpose here, however, is that most of the cases
in the dataset, whatever the types are, is basically caused by
financial insolvency state of the firm. Indeed, out of the total
16,526 cases, the most frequent one is suspension of bank transactions
(13,850). This liquidity problem (see appendix for details) is
considered as the main cause for other legal and private procedures of
bankruptcy.

Second, debt refers to the total sum of liabilities in the credit
(``right-hand'') side of balance-sheet. This quantity, when the firm
goes bankrupted, is relatively easy to measure in comparison
with other balance-sheet quantities such as total asset and equity%
\footnote{%
  Debt here is not net liabilities defined by the total liabilities
  minus the amount of assets not hypothecated, which is difficult to
  know.%
}. Since each item in debt has a creditor outside of the firm, so the
quantity is not basically hidden inside. Actual measurement of the
amount of debt is based either on legal documents or on investigation
of current/non-current liabilities in balance-sheet plus the amount of
bills discounted and transfers by endorsement, according to legal and
private procedure for bankruptcy, respectively.

We used the entire dataset in what follows.

\subsection{Distribution of debt}

\begin{figure}
  \centering
  \includegraphics[width=.7\columnwidth]{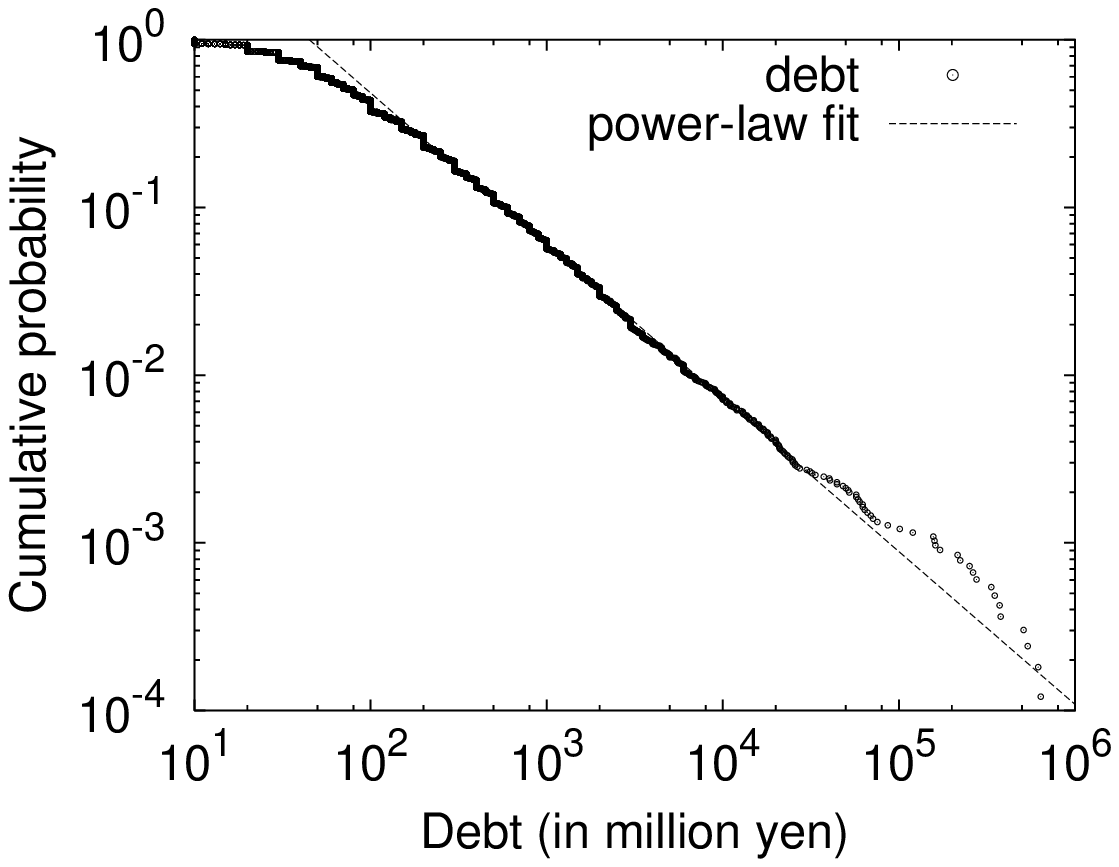}
  \caption[]{%
    Distribution of firms debt when bankrupted. The number of data
    amounts to 16526 firms bankrupted in the year of 1997.}
  \label{fig:debt}
\end{figure}

\figref{fig:debt} shows the cumulative probability distribution of
debt when the firm was bankrupted. For large debts exceeding $10^8$
yen, we can observe a power-law distribution over three orders of
magnitude or even more.  The probability that a given bankrupted firm
has debt equal to or greater than $x$, denoted by $P_>(x)$, obeys
\begin{equation}
  P_>(x) \propto x^{-\mu} ,
\label{eq:pareto}
\end{equation}
with a constant $\mu$, the so-called Pareto index for bankruptcy.
The index was estimated as $\mu=0.911\pm 0.008$ (by least-square-fit
for samples equally spaced in logarithm of rank in the power-law
region; error is 90\% level). This phenomenological finding is
what we call Zipf-law for debt of firms bankrupted. It was pointed out
in \cite{Aoyama00} that debt may obey Zipf-law, but unfortunately with
a quite limited number of data (corresponding to the tail more than
$10^{11}$ yen of debt).

\subsection{Life-time of bankrupted firms}

\begin{figure}
  \centering
  \includegraphics[width=.7\columnwidth]{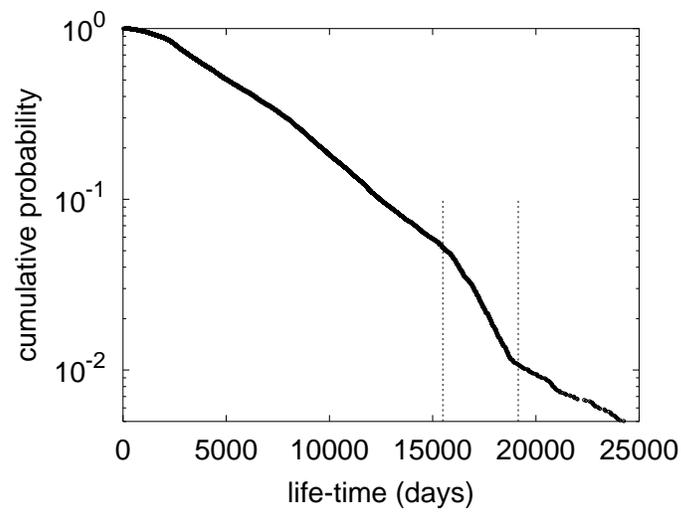}
  \caption[]{%
    Distribution of life-time for bankrupted firms. Time is given in
    units of days. It should be understood that the data may not
    contain the precise date of the firm's establishment, and can
    include one-month error. The two dotted lines show 10 years right
    after the world war II, during which the birth-rate of firms was
    extremely higher than other epochs.}
  \label{fig:life}
\end{figure}

No less interesting than how large a firm can fail is the problem of
how long a failed firm lived. \figref{fig:life} gives the cumulative
distribution of life-time $\tau$, $P_>(\tau)$, for the same set of
bankrupted firms. With this plot being semi-log, it is observed that
the entire distribution follows an exponential distribution. We also
see that the distribution has a sudden drop and kink in the shape. 

In order to quantify the shape of $P_>(\tau)$, let us introduce a
function defined by
\begin{equation}
  h(t)={-{dP_>(t)\over dt}\over P_>(t)} .
  \label{eq:haz}
\end{equation}
It is easily seen from the right-hand side expression of
\eqref{eq:haz} that for a sufficiently small time-interval $\Delta t$,
$h(t)\Delta t$ is the probability that the life-time $\tau$ satisfies
$t<\tau<t+\Delta t$ under the condition that $\tau\geq t$.  This
function has a similar notion as hazard rate function in probability
(see \cite{Ross96} for example). If $h(t)$ is constant $h_0$, then
$P_>(t)=\exp(-h_0 t)$. $h(t)$ has the physical dimension of inverse of
time.

\begin{figure}
  \centering
  \includegraphics[width=.95\columnwidth]{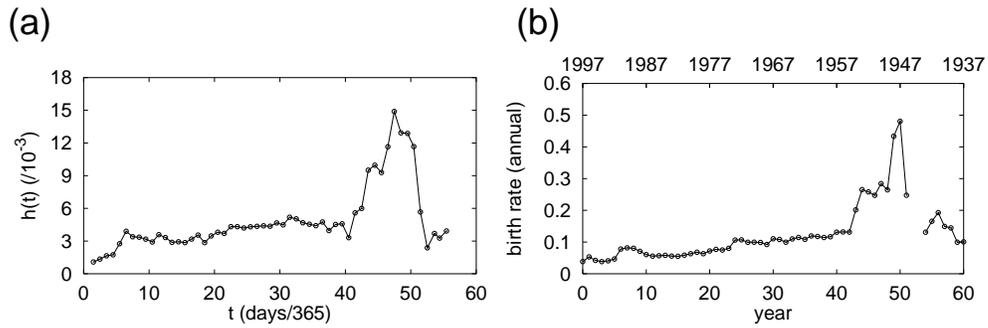}
  \caption[]{%
    (a) Estimation of $h(t)$ (per days) define in \eqref{eq:haz}. $t$
    is in units of 364 days and increases towards past. (b) Entry rate
    (per year) of new firms. Annual record of entry rate was made as
    described in \cite{ENTRY}. $t$ is given in years (lower tics) and
    in A.D. (upper). Data for 1944 and 1945 (the end of
    the World War II) are missing.%
  }
  \label{fig:life-brate}
\end{figure}

From $P_>(t)$ in \figref{fig:life}, we estimate the derivative in
\eqref{eq:haz} by doing moving-average of stepwise slopes over one
year. The resulting $h(t)$ is given in \figref{fig:life-brate}~(a).
$h(t)$ is nearly constant over 30 years, but takes large values for
about 10 years right after the World War II. In general, the number of
firm $t$ years old can be estimated by multiplying the number of firms
born $t$ years ago by the probability of survival of a firm to an age
of $t$ years. Therefore, if the process of survival is homogeneous in
time, the time-scale in $h(t)$ comes from entry rate of new firms.
\figref{fig:life-brate}~(b) is the historical record of {\it entry of
 firms\/} \cite{ENTRY}. One can observe that the 10 years period of
high $h(t)$ corresponds to the period of extremely high birth-rate
right after the world war II for about 10 years. This was brought
about by the dissolution of {\it zaibatsu\/} (corporate alliances)
monopoly under the government by U.S. General Headquarters (from 1945
to 1952). One can also note that $h(t)$ has the same order of
magnitude as the entry rate of firms by converting time from days to
years all over the history.

\subsection{Correlation between firm size and debt when bankrupted}

\begin{figure}
  \centering
  \includegraphics[width=.7\columnwidth]{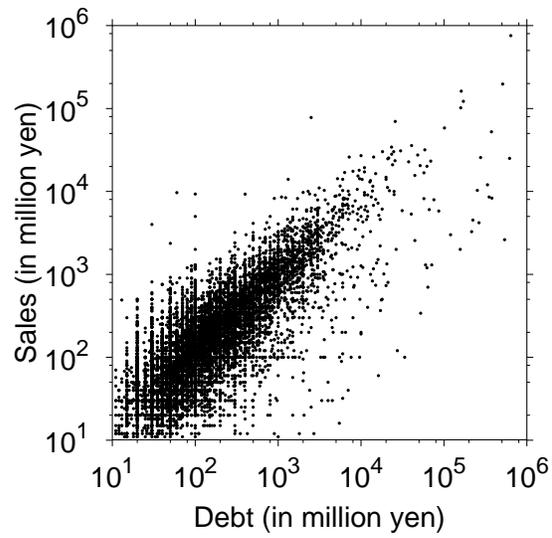}
  \caption[]{%
    Scatter plot for debt and sales (both in units of million yen).
    The values of sales refer to annual sales most recently available
    before bankruptcy.}
  \label{fig:d-s}
\end{figure}

Since it has been known%
\footnote{%
  But see also another paradigm \cite{Laherrere98}.%
} that firm size follows Zipf law
\cite{SIMON77,Amaral97,Okuyama99,Axtell01,Aoyama03}, let us take a
look at relation between firm size and firm debt right before
bankruptcy. \figref{fig:d-s} shows a clear correlation between them.
Here the firm size is measured by sales right before bankruptcy.  The
number of employees, as a proxy of firm size, gives similar result as
for sales. The bigger a firm is, the larger its debt is when failed.

\begin{figure}
  \centering
  \includegraphics[width=.8\columnwidth]{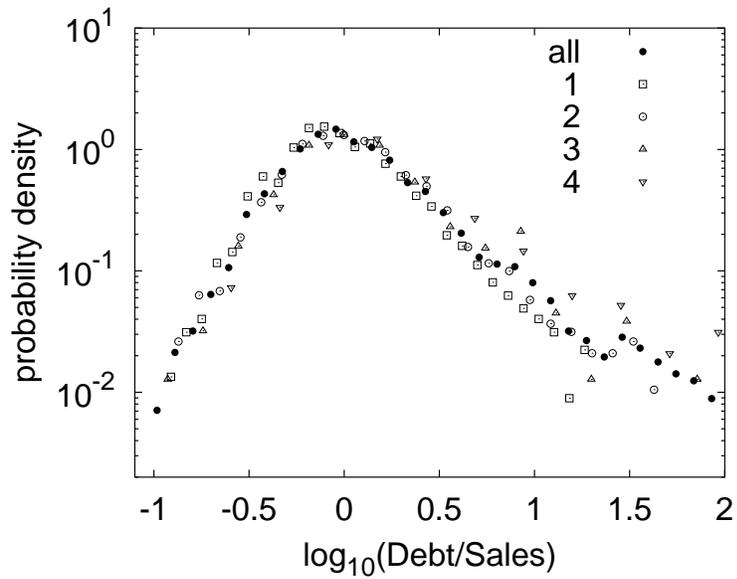}
  \caption[]{%
    Probability density $P(a)$ (filled dots) for the logarithmic ratio
    $a$ between debt $x$ and sales $s$, $a=\log_{10}(x/s)$. Also shown
    are probability densities $P(a|x)$ conditioned on the value of
    debt, $x\in[10^{8+0.4(n-1)},10^{8+0.4n}]$ (yen), where
    $n=1,\cdots,4$.}
  \label{fig:d-s.r}
\end{figure}

To quantify this relation, denoting sales of a bankrupted firm by $s$,
we use the ratio $A=x/s$ where debt $x$ and sales $s$ are given in the
same units of money. Introduce the logarithm of $A$ as $a=\log_{10}(A)$,
and we examined the probability density of $a$, $P(a)$, in
\figref{fig:d-s.r} (filled dots) by using all the data $x>10^8$ (yen),
which corresponds to the Pareto-Zipf range in \eqref{eq:pareto}.
It is found that the probability
density function $P(a)$ is exponential distribution with skewness,
i.e., with different slopes for $a>0$ and $a<0$. The side of $a>0$ is
fatter than the other side because a positive $a$ means that the firm
is nearer to the state of insolvency.

Furthermore, we show that the ratio does not depend statistically on
which range of debt is being observed. Preparing bins as
$x\in[10^{8+0.4(n-1)},10^{8+0.4n}]$ yen ($n=1,\cdots,4$), we can
examine the conditional probability density, $P(a|x)$, for each bin of
$x$. \figref{fig:d-s.r} shows that statistical dependence of the ratio
on the value of debt is very weak. This means that the relation
between firm size and debt is multiplicative with multiplying factor
$A$ drawn from the distribution given in \figref{fig:d-s.r}. If the
statistical independence between $A$ and $s$ holds over the entire
range of the variables, the probability density for $\log x$ is equal
to the convolution of the one for $a=\log A$ and the one for $\log s$.
If $s$ has a power-law distribution with a Pareto index, then $x$ will
have a power-law with the same Pareto index. Though being a
non-rigorous argument because of the breakdown of power-law under some
threshold and of the statistical independence, this give us a basic
idea that the Zipf law for firm size and that for firm debt when
bankrupted are closely related to one another.

\section{Balance-sheet dynamics of firm}

\subsection{What is a firm?}

In order to understand how firms die, let us consider the activity of
a firm from the viewpoint of money flow. 

A firm makes activity by producing goods or services in anticipation
to be demanded by consumers, and to yield profit. For the firm to
realize the production and selling of goods and services, it must
usually invest more money than what it actually possesses with a fund
by the firm's owners or stockholders. The investment is done not only
for cost of sales, expenses of selling, etc. but also for additional
facility, production line, employment etc.

A firm's activity has three facets in general:
\begin{enumerate}
\item finance\\
  borrowing money from banks, market investors
\item investment\\
  spending money in anticipation of future return of profits
\item collecting\\
  earning profits from the sales of good and services
\end{enumerate}
In other words, these three facets imply (1) the presence of
creditor (borrower), (2) that investment can be a source of risk,
because (3) return of profits has uncertainty in realization.

\begin{figure}
  \centering
  \includegraphics[width=.7\columnwidth]{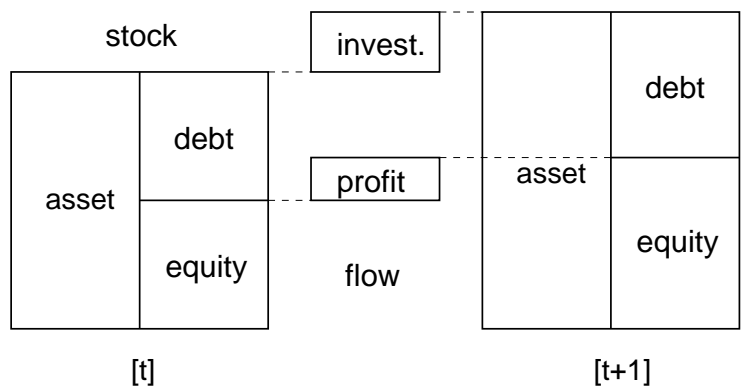}
  \caption[]{%
    Balance sheet dynamics. From time $t$ to time $t+1$ the stock of
    asset composed of debt and equity is subject to change according
    to the flow of realized profit and investment during the period.
  }
  \label{fig:bsdyn}
\end{figure}

The above description can be formulated in terms of stock and flow of
money. {\it Flow\/} is inflow of money in a period of time, and {\it
 stock\/} is the accumulation of flow over time. Stock of a firm is
its balance sheet which consists of asset, debt and equity (see
\figref{fig:bsdyn}). Here asset refers to all the assets possessed by
the firm which are expected to yield profits. The asset can be
classified into equity and debt according to how it is financed,
either by stockholders or by loan from banks and market, respectively.

Temporal change of balance-sheet is determined by flow. One source of
flow is profit which is basically given by sales minus cost of sales,
finance, and others. The other flow is investment which is determined
by the firm's decision based on anticipation of profits. Since the
investment is covered by additional debt, it also depends on
creditor's activity.

Balance sheet dynamics is the temporal change of asset and debt. Debt
is necessary for making profits, but is a source of risk since it can
diminish equity and can dominate the balance sheet. Financial
fragility can be measured either by flow variable such as insolvency
(the ratio between debt commitments to profits) or by stock variable
such as liquidity (the ratio between debt and capital). Persistent
state of insolvency results in liquidity problem. Actually the most
cases of bankruptcies (see Appendix) are due to the problem of
liquidity.

\subsection{Aggregate dynamics of creditor and debtors}

Recently a model for the dynamics of balance sheets of a bank and
firms are proposed in \cite{Gallegati03}, which is based on their
previous model for aggregate dynamics of firms \cite{Gallegati00}.
This model of creditor-debtor dynamics has two elements; competition
between firms by their sizes, and debtor-creditor relation between a
bank and each firm.

In this section, let us briefly recapitulate the model (see
\cite{Gallegati03} for details) and show by simulation that our
phenomenological findings about bankruptcy can be reproduced in the
model.

A firm $i$ has, at time $t-1$, asset $K^i_{t-1}$, debt $L^i_{t-1}$ and
equity $A^i_{t-1}$ which satisfy the identity:
\begin{equation}
  K^i_{t-1}=L^i_{t-1}+A^i_{t-1}.
  \label{eq:firm_id}
\end{equation}
$i=1,\cdots,n$ where $n$ is the number of firms. In the period of
time from time $t-1$ to $t$, the firm makes the following activity
with $K^i_t$ to be determined below. It produces an output $\phi K^i_t$
($0<\phi<1$), which is to yield revenue with uncertainty. At the same
time, the firms has financial cost, comprised of dividend for equity
and interest rate for debt. The profit in the current period is
\begin{equation}
  \pi^i_t=u^i_t\phi K^i_t-r^i_t K^i_t ,
  \label{eq:firm_p}
\end{equation}
where $r^i_t$ is interest-rate for the firm $i$. Here $u^i_t$ is a
stochastic variable representing the uncertainty in making profit. In
the second term, the ratio of dividend to equity is assumed to be
equal to the interest rate.

Supposing $r^i_t$ is a parameter, the firm plans the amount of
investment during the current period. Since the profit $\pi^i_t$ is
proportional to the total asset $K^i_t$, the firm could increase its
size by investment. But the bigger the size is, the larger the risk of
insolvency or bankruptcy is, in the presence of uncertainty for
profit. As a simplest form, the objective function for determining the
investment is assumed to have a quadratic term of $K^i_t$. Namely, the
maximization of the expectation:
\begin{equation}
  E[ \pi^i_t - c(K^i_t)^2 ]
  \label{eq:firm_opt}
\end{equation}
yields an optimal value of $K^{i*}_t$. Then the investment in the
current term is
\begin{equation}
  I^i_t=K^{i*}_t-K^i_{t-1}
  \label{eq:firm_inv}
\end{equation}
for the realization of which the firm demand loan of the amount
\begin{equation}
  L^{i(d)}_t=L^i_{t-1}-\pi^i_{t-1}+I^i_t .
  \label{eq:firm_debt}
\end{equation}
In this model the demand is assumed to be always fulfilled by bank
with an appropriate interest-rate. Therefore balance sheet updates by
$K^i_t=K^{i*}_t$, $L^i_t=L^{i(d)}_t$, and $A^i_t=K^i_t-L^i_t$, once
the interest rate $r^i_t$ is determined as shortly given.

Bank is monopolistic and responsible for all the loans to firms. Its
balance sheet is composed of loan $L_{t-1}$ as asset, deposit
$D_{t-1}$ as debt, and equity $E_{t-1}$ of itself:
\begin{equation}
  L_{t-1}=D_{t-1}+E_{t-1} .
  \label{eq:bank_id}
\end{equation}
The total supply of loans is usually determined by the ratio of asset
to equity, so is assumed here as
\begin{equation}
  L_{t-1}={1\over\alpha} E_{t-1}
  \label{eq:bank_roe}
\end{equation}
with a parameter $0<\alpha<1$. The supply of loan to each firm is
assumed to be based of collateral of firm asset.
\begin{equation}
  L^{i(s)}_t=L_t{K^i_{t-1}\over\sum_{j=1}^n K^j_{t-1}}
  \label{eq:bank_loan}
\end{equation}
This is where competition between firms is in effect. Then the
interest-rate for each firm is determined by
$L^{i(s)}_t=L^{i(d)}_t$. The bank's profit is revenue from interests
minus the financial cost of the bank
\begin{equation}
  \Pi^i_{t}=\sum_{i=1}^n r^i_t L^i_t - (\bar{r}_t-\omega) L_t
  \label{eq:bank_p}
\end{equation}
where
\begin{equation}
  \bar{r}_t=\sum_{i=1}^n{L^i_t\over L_{t-1}}
  \label{eq:avg_r}
\end{equation}
and $\omega$ is a parameter for profit markup.

The bank faces uncertainty of firms bankruptcy. A firm goes into
bankruptcy when its equity becomes negative. Bankrupted firms are
replaced with new ones with initial and same balance sheet. The update
for the bank' is thus given by
\begin{equation}
  E_t=E_{t-1}+\Pi_t-\sum_{i=1}^n B^i_{t-1}
  \label{eq:bank_up}
\end{equation}
where $B^i_{t-1}$ is $-A^i_{t-1}$ for every firm $i$ bankrupted
($A^i_{t-1}<0$), otherwise 0.

The dynamics of this system was investigated by simulation in
\cite{Gallegati03}. It was found that the model has, in a stationary
phase, skewed distribution of firms sizes with power-law, Laplace
distributed growth rates, and that small idiosyncratic shocks generate
large aggregate fluctuations. We used the same set of parameters:
$c=1$, $\phi=0.1$, $\alpha=0.08$, $\omega=0.002$ except the number of
firms $n=10^5$ (much larger than $10^2$ in original). The stochastic
variable $u^i_t$ obeys a uniform distribution with support (0,2).  The
results of simulation are robust with other set of parameters, and
with the number of firms.

\begin{figure}
  \centering
  \includegraphics[width=.95\columnwidth]{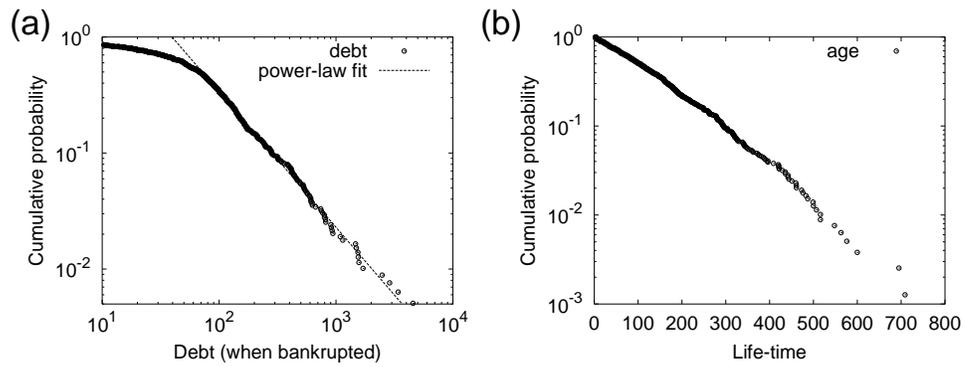}
  \caption[]{%
    (a) Cumulative probability distribution of debts bankrupted in a
    time step during a stationary phase. (b) Cumulative probability
    distribution of life-time for the same family of bankrupted firms.
    The horizontal value is in arbitrary units for both plots.
  }
  \label{fig:sim}
\end{figure}

Our new findings from the simulation is given in \figref{fig:sim}. We
can observe that the distributions of debt and life-time for
bankrupted firms have the same qualitative results as in the real
data. These results are also robust for different parameters.

\section{Summary and Discussion}

Firms growth and failure are the two sides of the same coin. In this
paper, we have uncovered the phenomenology of firms bankruptcy by
using an exhaustive list of Japanese bankruptcies in 1997. Namely,
for high-debt regime
\begin{enumerate}
\item debt of bankrupted firms follows Zipf distribution
\item debt and size of such firms are related by multiplicative factor
  independent of the amount of debt
\item bankrupted firms have life-time whose distribution is
  exponential but correlates with entry rate of new firms
\end{enumerate}

The origin of Zipf law in firms bankruptcy might be well captured in
the model \cite{Gallegati03}, which was shown to reproduce the
phenomenological properties in bankruptcy. Indeed the model shares the
aspect of self-organized criticality \cite{Bak96} (see also
\cite{Sornette00}) as discussed briefly in what follows.

Failure of a firm can have tremendous influence to economic activities
in macroscopic scale. This is because the default of a single firm can
propagate its effect through a network of creditor-debtor
relationship. In the model, a bankrupted firm leaves the market
without paying back its debt and debt commitments to bank. Then the
banking system suffers from a capital loss, resulting in the shrink of
credit supply. Such shrink accompanied with rise of interest rates is
likely to strengthen the competition between firms for seek of more
credit and then to deteriorate the firms balance-sheets. But this
results in the growing probability of bankruptcy. This is a ``domino
effect''. The model also shows the opposite case of financially
infragile or sound state of firms and bank which leads growth of the
total credit and so that of total production (since the system is
assumed to have enough money supply when necessary). 

During this process of debtor-creditor dynamics, all the firms are
competing in their sizes having the risk of bankruptcies. The risk is
shared by the bank who attempts to make profit by interest rate. Note
that the model has essentially no fine-tuning parameter for having
$\mu=1$ in \eqref{eq:pareto} and the Zipf law in firms size. In this
respect, the model differs from the traditional approach of stochastic
processes for firms growth and failure (see \cite{Steindl65} for
mid-60's nice review) including entry-exit processes, collective risk
theory, etc. Rather it is the heterogeneous interacting and competing
agents that brings about critical state in the aggregate dynamics of
firms (see \cite{Aoki02} for this new direction).

\begin{ack}
  We acknowledge Tokyo Shoko Research, Ltd. for providing high-quality
  data. We are grateful to thank Mauro Gallegati, Gianfranco Giulioni
  and Nozomi Kichiji for providing information about their model,
  Hideaki Aoyama and Wataru Souma for discussion, Masanao Aoki for
  guiding the author to the model, and Katsunori Shimohara for
  encouragement. The author is indebted to Mauro Gallegati for
  invaluable comments and suggestions. Supported in part by grants
  from the Telecommunications Advancement Organization in Japan and
  from Japan Association for Cultural Exchange. 
\end{ack}


\clearpage
\clearpage
\clearpage
\clearpage
\clearpage
\clearpage
\clearpage


\clearpage
\appendix
\section{Bankruptcy data}

Bankruptcy or business failure is not a legal term, but should be
understood as a commonly used term to describe a state of critical
financial insolvency of a debtor.  A typical case of bankruptcy is a
suspension of banking transactions (see below). What are also termed
bankruptcy are: filing to the court for bankruptcy proceedings or a
corporate rehabilitation procedure, or informing the aggravation of
financial condition of the debtor and delegating creditors to handle
it.

Types of bankruptcy are legally divided into legal and private
procedures. The former procedure has purpose for rehabilitation or
liquidation. Rehabilitation is done by either of Corporate
Reorganization Law, Civil Rehabilitation Law (since 2000), Corporate
Arrangement under the Commercial Code, or Composition Law (abandoned
in 2000). Liquidation is done by Special Liquidation or by
``Bankruptcy''. Private procedure, on the other hand, refers to either
Suspension of Bank Transactions or Internal Arrangement. The number of
each bankruptcy types in 1997 is as follows:
\begin{center}
{\small
\begin{tabular}{|l|r|}
  \hline
  bankruptcy type & number \\
  \hline
  \multicolumn{2}{|l|}{private procedure} \\
  \hspace{3ex} Suspension of Bank Transactions & 13850 \\
  \hspace{3ex} Internal Arrangement & 501 \\
  \hline
  \multicolumn{2}{|l|}{legal procedure for rehabilitation} \\
  \hspace{3ex} Composition Law & 195 \\
  \hspace{3ex} Corporate Reorganization & 23 \\
  \hspace{3ex} Corporate Arrangement & 12 \\
  \hline
  \multicolumn{2}{|l|}{legal procedure for liquidation} \\
  \hspace{3ex} ``Bankruptcy'' & 1877 \\
  \hspace{3ex} Special Liquidation & 68 \\
  \hline
  \hline
  total & 16526 \\
  \hline
\end{tabular}
}
\end{center}

The most frequent case is suspension of bank transactions.  Concerning
with the bill and check clearings, clearing houses practice a
punishment system against the entity who committed dishonored bills or
checks in order to maintain the public credibility. The bills or
checks which a debtor fails to settle within the term shall be
``dishonored''. The second dishonor within six months after the first
one shall result in publishing the debtor's name on the "the report on
suspension of banking transactions."

Then the debtor shall be barred from transactions with any member
financial institutions of the same clearinghouse for two years from
the date of suspension, being prohibited from opening trade accounts,
drawing bills or checks, and receiving loans. In actuality, this is
equivalent to bankruptcy of the firm.

In summary, we regard this case as {\it liquidity problem\/}, and also
consider that other legal and private procedures are basically due to
the same problem (debt growth).

\end{document}